\documentclass[final,twocolumn]{elsarticle}
\usepackage{amssymb}

\journal{Physics Letters A}

\begin{document}

\begin{frontmatter}

\title{Critical and multicritical behavior in the Ising-Heisenberg universality class}

\author{A.O. Sorokin}
\ead{aosorokin@gmail.com}

\address{Petersburg Nuclear Physics Institute, NRC Kurchatov Institute, 188300 Orlova Roscha, Gatchina, Russia}

\begin{abstract}
Critical behavior of three-dimensional classical frustrated antiferromagnets with a collinear spin ordering and with an additional twofold degeneracy of the ground state is studied. We consider two lattice models, whose continuous limit describes a single phase transition with a symmetry class differing from the class of non-frustrated magnets as well as from the classes of magnets with non-collinear spin ordering. A symmetry breaking is described by a pair of independent order parameters, which are similar to order parameters of the Ising and O(N) models correspondingly. Using the renormalization group method, it is shown that a transition is of first order for non-Ising spins. For Ising spins, a second order phase transition from the universality class of the O(2) model may be observed. The lattice models are considered by Monte Carlo simulations based on the Wang-Landau algorithm. The models are a ferromagnet on a body-centered cubic lattice with the additional antiferromagnetic exchange interaction between next-nearest-neighbor spins and an antiferromagnet on a simple cubic lattice with the additional interaction in layers. We consider the cases N=1,2,3 and in all of them find a first-order transition. For the N=1 case we exclude possibilities of the second order or pseudo-first order of a transition. An almost second order transition for large N is also discussed.
\end{abstract}

\begin{keyword}
Frustrated magnets  \sep
Phase transitions \sep
Order from disorder \sep
Monte Carlo simulations \sep
Wang–Landau algorithm \sep
Renormalization group \sep
\end{keyword}

\end{frontmatter}

\section{Introduction}

A complicated tensor structure of an order parameter differing from the case of the usual vector $O(N)$ model is realized in many physical systems. During the last several decades, such models are investigated intensively (see \cite{Vicari02} for a review). In context of magnetic systems, one of the most interesting model is the matrix $O(N)\otimes O(M)$ model describing frustrated magnets with non-collinear (planar $M=2$ or non-planar $M=3$) spin ordering.  Besides canted and sinusoidal phases of magnets, this model has been also discussed in the context of other systems of condesed matter physics such as superfluid $^3$He \cite{Jones76,Bailin77,Sokolov79,Sokolov81}, some types of superconductors \cite{Volovik85,Kumar87}, Josephson junction arrays in a magnetic field at zero temperature \cite{Yosefin85,Granato90}, etc.

Another important model with a complicated structure of an order parameter is the $O(N_1)\oplus O(N_2)$ model comprises two interacting vector models. It describes a multicritical point and an intersection (or junction) of two critical lines corresponding to different vector order parameters. Such a multicritical point arises in antiferromagnets in an external field \cite{Fisher74,Aharony03,Kosterlitz74,Pokrovsky75,Kosterlitz76}, the $O(5)$ theory of high temperature superconductors \cite{Zhang97,Hu01,Hasenbusch05}, etc.

In the present work, we consider another symmetry breaking scenario realized in certain models of frustrated antiferromagnets. Namely, we are interested in the so-called Ising-Heisenberg symmetry class describing a single phase transition with breaking of $\mathbb{Z}_2\otimes SO(N)/SO(N-1)$ symmetry. According to the symmetry, this class corresponds to magnets with a collinear spin ordering and with an additional twofold degeneracy of the ground state. Two such models have been considered for quantum spins. The first model is a ferromagnet on a body-centered cubic lattice with the additional antiferromagnetic exchange interaction between next-nearest-neighbor spins \cite{Schmidt02,Oitmaa06,Majumdar09,Pantic14,Farnell16,Mi16,Mi17}. And the second one is a antiferromagnet on a simple cubic lattice with the additional interaction in layers \cite{Schmalfuss06,Nunes10,Holt11,Majumdar11,Rojas11,Isaev12,Fan14}. We investigate the critical behavior of these two models in the case of classical spins using Monte Carlo simulations.

To confirm the numerical results and to generalize them for the whole Ising-Heisenberg universality class, we consider the continuous limit of the lattice models using renormalization group (RG) approach. Fortunately, the corresponding Ginzburg-Landau-Wilson (GLW) functional is closely related to the well-studied $O(N)\otimes O(M)$ and $O(N_1)\oplus O(N_2)$ models, that allows to base our RG calculations on the results obtained for these models.

The static critical phenomena in the $O(N)\otimes O(M)$ model are described by the GLW functional \cite{Garel76,Bak76,Brazovsky76,Dzyaloshinsky77,Barak82,Kawamura88}
\begin{eqnarray}
    F=&&\int d^3x\left((\partial_\mu\phi)^2+(\partial_\mu\psi)^2+r(\phi^2+\psi^2)+
    \right.\nonumber\\
    &&\left.
    u\left(\phi^2+\psi^2\right)^2+2w\left((\phi\psi)^2-\phi^2\psi^2\right)\right),
    \label{ONxOM}
\end{eqnarray}
with $w>0$, and $\phi$, $\psi$ are $N$-component vector fields. This model are investigated in the framework of several approaches as the $4-\epsilon$ expansion \cite{Sokolov95,Pelissetto01,Calabrese04}, perturbative \cite{Sokolov94,Pelissetto01-2,Pelissetto01-3,Calabrese02,Calabrese03,Holovach04,Delamotte08,Delamotte10} and non-perturbative RG \cite{Zumbach93,Zumbach94,Zumbach94-2,Delamotte00,Delamotte03}, $1/N$ expansion \cite{Pelissetto01,Gracey02,Gracey02-2}, and numerical studies of corresponding lattice models. For a review including numerical and experimental results, see \cite{Delamotte04}.

The critical behavior at a multicritical point can be studied by the GLW functional
\begin{eqnarray}
    F=&&\int d^3x\left((\partial_\mu\phi)^2+(\partial_\mu\psi)^2+r(\phi^2+\psi^2)+\right.\nonumber\\
    &&\left.u_1\phi^4+u_2\psi^4+2v\phi^2\psi^2\right).
    \label{On+Om}
\end{eqnarray}
This model has been investigated within the $4-\epsilon$ expansion \cite{Kosterlitz74,Pokrovsky75,Kosterlitz76,Pelissetto03}, perturbative $d=3$ \cite{Prudnikov98,Folk08} and  non-perturbative RG (NPRG) \cite{Eichhorn13}. In addition, the NPRG approach has been used in \cite{Wetterich95,Wetterich96} to study the case $N_1=N_2=1$ of the model (\ref{On+Om}), where $O(1)\equiv \mathbb{Z}_2$ corresponds to a symmetry of the Ising model. This case is of special interest in statistical physics in the context of the Ashkin-Teller model \cite{Ashkin43,Fan72} and its critical line corresponding to a singular transition in both Ising order parameters. In three dimensions, the Ashkin-Teller model has been studied numerically in \cite{Kadanoff80,Oliveira89,Arnold97,Musial04,Musial06}. Intensive numerical studies have been also performed in recent works \cite{Landau78,Selke11,Hasenbusch11,Selke13,Landau14} for the multicritical point of a $N=3$ antiferromagnet in a magnetic field. This point is described by the $\mathbb{Z}_2\oplus O(2)$ model. The fermionic extension of the later model has been studied in a context of a critical point between semimetallic and insulating phases in graphene \cite{Herbut06,Herbut09,Herbut09-2,Roy11}.

As we show in this paper, the critical behavior in the Ising-Heisenberg class is described by the GLW functional
\begin{eqnarray}
    F=&&\int d^3x\left((\partial_\mu\phi)^2+(\partial_\mu\psi)^2+r(\phi^2+\psi^2)+
    \right.\nonumber\\
    &&\left.
    u\left(\phi^4+\psi^4\right)+2v\phi^2\psi^2+2w(\phi\psi)^2\right),
    \label{Z2xON}
\end{eqnarray}
with some positive constants $u$ and $v$, and negative $w$. The ground state of the model (\ref{Z2xON}) strongly depends on a sign of the coupling constant $w$. When $w<0$, the vectors $\phi$ and $\psi$ tend to be parallel. This is the case corresponding to the lattice models discussed in the present work. The case $w=0$ returns us to the model (\ref{On+Om}) with $u_1=u_2$, and $w>0$ describes magnetic systems with a planar spin ordering (\ref{ONxOM}). The last case has been also considered in the more general model in \cite{Pismak09}. These relations between the three models are very useful, in particular, we can extrapolate the $4-\varepsilon$ expansion up to the five-loop order using the works \cite{Pelissetto01,Calabrese04} and \cite{Pelissetto03}.

At $N=1$, the models (\ref{On+Om}) and (\ref{Z2xON}) are equivalent. This case describes two interacting Ising models. Previous investigations cited above predict that the critical behavior of a single phase transition in both Ising order parameters can be either from the universality class of the $O(2)$ model or a phase transition is of first order dependently on details of a microscopic (lattice) model. Also, two marginal critical behaviors are realized in this case: tricritical one, and Ising one corresponding to decoupled Ising models. The last case corresponds to the large second exchange limit for both considered lattice models. Using Monte Carlo simulations, we show that in the lattice models a phase transition is of first-order. This is consistent with the results of the $4-\varepsilon$ \cite{Banavar79} and temperature \cite{Oitmaa79,Ferer83} expansions for the $N=1$ model on a body-centered cubic lattice, but contradicts to the recent Monte Carlo simulations \cite{Murtazaev15,Murtazaev15-2} where a second-order phase transition has been found with a novel set of the critical exponents. Studies of a single phase transition in the Ashkin-Teller model confirm our result with a first-order transition \cite{Arnold97}.

In the case $N=2$, $\mathbb{Z}_2\otimes SO(2)$ symmetry is broken. Noteworthy, the same symmetry is broken in XY antiferromagnets with a planar ordering, such as helimagnets (see \cite{Sorokin14} and references therein).  So the case $N=2$ of the model (\ref{Z2xON}) is equivalent to the case $N=2$ of the model (\ref{ONxOM}), although these cases correspond to the different regions ($w<0$ and $w>0$) of the coupling constant space of the model (\ref{Z2xON}). Moreover, Monte Carlo simulations of the lattice models \cite{Sorokin14-2} show that a phase transition is of weak first order and pseudo-scaling exponents are close exponents of a XY helimagnet \cite{Sorokin14}.

The cases $N>2$ are novel universality classes. Since a second-order phase transition is not observed in $N=2,3$ magnets with non-collinear spin orderings \cite{Delamotte04}, the Ising-Heisenberg class becomes the main candidate for a searching novel types of the critical behavior \cite{Sorokin12}. However, we find that a phase transition is of weak first order. The same results have been obtained for similar model from the Ising-Heisenberg universality class in \cite{Batista09,Batista10,Batista11}.

This paper is organized as follows. In section \ref{sect2}, we start from the lattice models and, acting in a standard manner, obtain the GLW functional (\ref{Z2xON}).  Possible scenarios of symmetry breaking in the $\mathbb{Z}_2\otimes O(N)$ models as well as cases where the symmetry of the functional enlarges are discussed in section \ref{sect3}. And in section \ref{sect4} the RG analysis of this model is performed. It turns out that a significant part of information about fixed points and critical behavior at them is reduced to the solution of models (\ref{On+Om}) and (\ref{ONxOM}). Although new additional non-trivial fixed points are present in the model (\ref{Z2xON}), they do not respond to $\mathbb{Z}_2\otimes SO(N)/SO(N-1)$ symmetry breaking. So we find that a phase transition from the Ising-Heisenberg universality class is of first order for $N\geq2$. The exception is the case $N=1$ corresponding to Ising spins. In this case, a transition can be of second order from the $O(2)$ universality class. Section \ref{sect5} is devoted to the results of Monte Carlo simulations for the cases $N=1,\,2$ and 3. The case $N=1$ is the most intriguing, where we discuss possibilities of the second order and pseudo-first order of a transition.

\section{Lattice models and their continuous limit}
\label{sect2}

\begin{figure}[t]
\center
\includegraphics[scale=0.32]{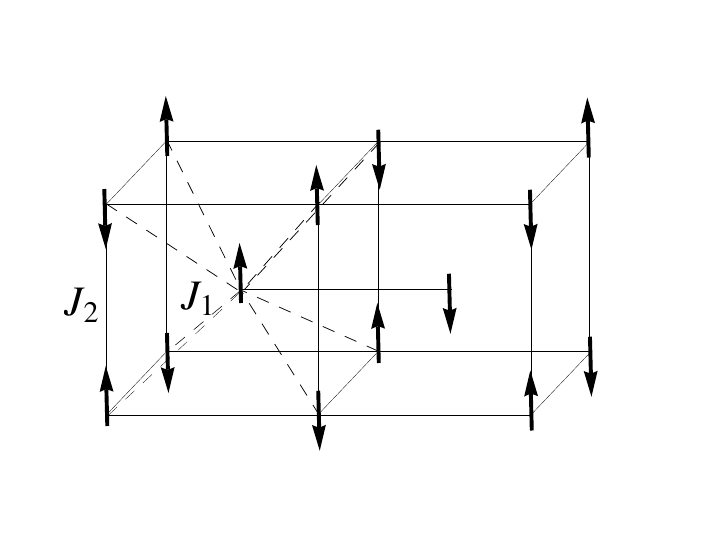}
\includegraphics[scale=0.32]{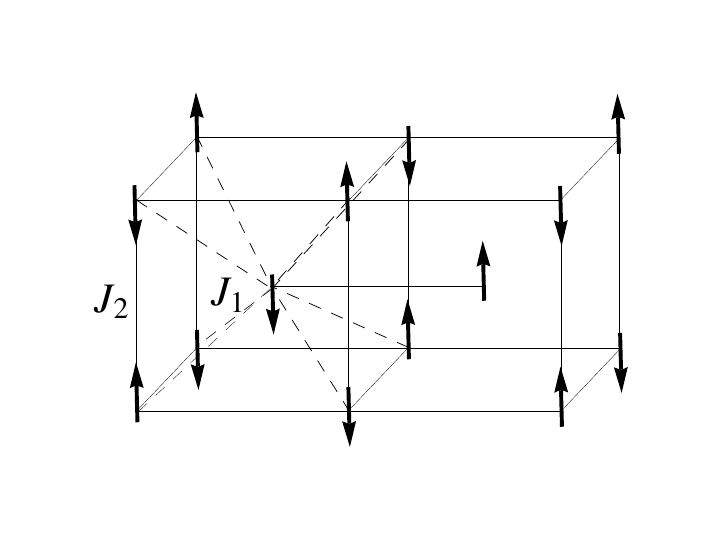}
\caption{\label{fig1}Two non-equivalent ground states of the J$_1$-J$_2$ model on a body-centered cubic lattice, which can not be reduced to each other through global spin rotations.}
\end{figure}

A scenario of $\mathbb{Z}_2\otimes SO(N)/SO(N-1)$ symmetry breaking is realized in frustrated antiferromagnets with a collinear spin ordering and twofold degeneracy of the ground state. Among three-dimensional model, such a structure of the ground state is present in a antiferromagnet on a body-centered cubic lattice with the Hamiltonian
\begin{equation}
    H=J_1\sum_{ij}\mathbf{S}_i\cdot\mathbf{S}_j+J_2\sum_{kl}\mathbf{S}_k\cdot\mathbf{S}_l,
    \label{model1}
\end{equation}
where the sum $ij$ runs over pairs of nearest-neighbor spins, and the sum $kl$ runs over pairs of next-nearest-neighbor spins. A spin $\mathbf{S}$ is a classical $N$-component unit vector, $J_1,\,J_2>0$. At $J_2<2J_1/3$, the ground state is two embedded to each other ferromagnetic sublattices interacting antiferromagnetically, without a frustration. But at $J_2>2J_1/3$, sublattices become antiferromagnetic, and the ground state acquires the desired structure (see fig. \ref{fig1}). Strictly speaking, in the absence of
thermal or quantum fluctuations, the relative spin orientation between two sublattices are not determined. It means that the ground state is infinitely degenerated for $N\geq2$, but this degeneracy is lifted by fluctuations, so only two non-equivalent configurations (fig. \ref{fig1}) survive \cite{Shender82}. Such a phenomenon is known as "order from disorder".

\begin{figure}[t]
\center
\includegraphics[scale=0.32]{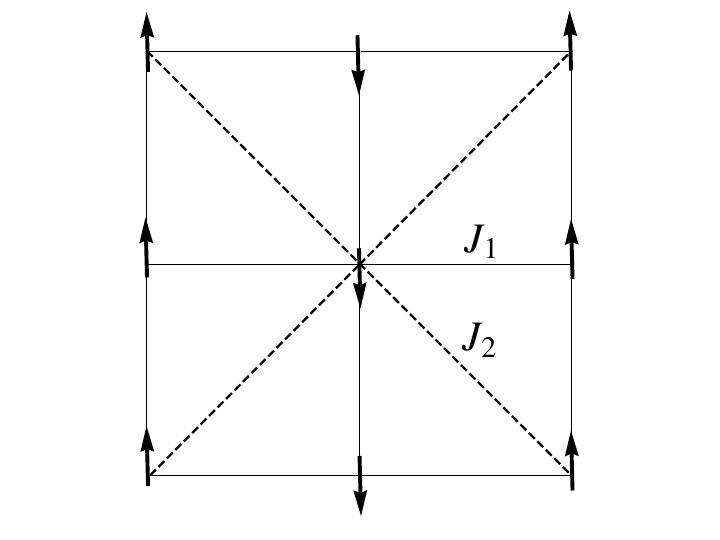}
\includegraphics[scale=0.32]{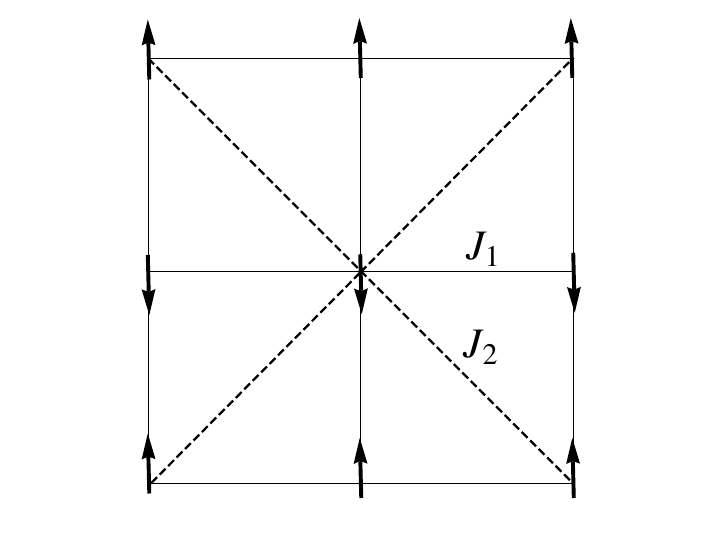}
\caption{\label{fig2}Two non-equivalent ground states of the stacked-J$_1$-J$_2$ model on a simple cubic lattice, which can not be reduced to each other through global spin rotations.}
\end{figure}

Another model, discussed also in two dimensions \cite{Henley}, is the stacked two-exchange model (stacked-J$_1$-J$_2$ model) on a simple cubic lattice with the Hamiltonian
\begin{equation}
    H=-J_1\sum_{ij}\mathbf{S}_i\cdot\mathbf{S}_j+J_2\sum_{kl}\mathbf{S}_k\cdot\mathbf{S}_l,
    \label{model2}
\end{equation}
where the sum $ij$ runs over pairs of nearest-neighbor spins, and the sum $kl$ enumerates pairs of next-nearest-neighbor spins in layers (see fig. \ref{fig2}). At $J_2<J_1/2$, the ground state is the ferromagnetic order. At $J_2>J_1/2$, the ground state is one of two spin configurations with the wave-vectors $\mathbf{q}=(\pi,0,0)$ or $\mathbf{q}=(0,\pi,0)$. This model is convenient and more expository for the derivation of the GLW functional and the continuous limit of these lattice models.

Acting in a standard manner, we obtain an equivalent model without the constrained field $\mathbf{S}$, but with an additional potential and coupling constants defining the length of the new field $\varphi$. For the GLW-approach, it is reasonably to hold just up to quartic terms in field of this additional potential $U(|\varphi|)=m\varphi^2+\lambda\varphi^4$. Further, to obtain the continuum limit of the lattice model, one should expand the expression in the vicinity of both minimum $(\pi,0,0)$ è $(0,\pi,0)$. We introduce the fields
\begin{eqnarray}
    \phi=\varphi|_{\mathbf{q}\simeq(\pi,0,0)}+\varphi|_{\mathbf{q}\simeq(0,\pi,0)},\nonumber\\
    \psi=\varphi|_{\mathbf{q}\simeq(\pi,0,0)}-\varphi|_{\mathbf{q}\simeq(0,\pi,0)},
\end{eqnarray}
so that its parallelism corresponds to the minimum $(\pi,0,0)$, and another minimum $(0,\pi,0)$ corresponds to antiparallel fields. Finally, we obtain the GLW functional (\ref{Z2xON}) corresponding to the starting lattice models.

\section{Symmetry and mean-field analysis}
\label{sect3}

As we have discussed above, the symmetry and the ground state of the model (\ref{Z2xON}) strongly depends on a sign of the coupling constant $w$. Let's consider all possibilities.

\subsection{$w<0$}

The extremum of the free energy functional (\ref{Z2xON}) in the ordered phase ($r<0$) attains on a homogeneous configuration satisfying to the conditions
\begin{equation}
    \phi_0^2=\psi_0^2=\frac{-r}{2(u+v+w)}\equiv \kappa^2,\quad \phi_0\parallel\psi_0.
    \label{absphi}
\end{equation}
This extremum is the global minimum in the stability region
\begin{equation}
    u>0, \quad w<0, \quad u+v+w>0, \quad u-v-w>0.
\end{equation}
For symmetry analysis, it is convenient to represent the order parameter as a $2\times N$ matrix $\Phi=\{\phi,\psi\}$. The functional (\ref{Z2xON}) is invariant under the left action of orthogonal matrices on the order parameter $\Phi\to T\Phi$, where $T\in O(N)$. Also, it is invariant under the right action of $2\times2$ orthogonal matrices corresponding to the three discrete $\mathbb{Z}_2$ symmetry generators
\begin{equation}
\left(
           \begin{array}{cc}
             0 & 1 \\
             1 & 0 \\
           \end{array}
         \right), \quad
\left(
           \begin{array}{cc}
             -1 & 0 \\
             0 & -1 \\
           \end{array}
         \right), \quad
\left(
           \begin{array}{cc}
             1 & 0 \\
             0 & -1 \\
           \end{array}
         \right),
\end{equation}
and their combinations. One has just eight such matrices, including the unit one, and they are elements of the group $(\mathbb{Z}_2\otimes\mathbb{Z}_2\otimes\mathbb{Z}_2)_R\subset O(2)_R$. The first subgroup replaces the vectors $\phi$ and $\psi$ between themselves $\phi\leftrightarrow\psi$. The second and third ones change a direction of two or one vectors to an opposite $\phi\to-\phi$ and/or $\psi\to-\psi$. Using all of these right as well as left acting symmetries, one can read the ground state as $\phi_0=\psi_0=(\kappa,0,\ldots,0)$. The spontaneously broken symmetry is
\begin{equation}
    \frac{O(N)_L\otimes(\mathbb{Z}_2\otimes\mathbb{Z}_2\otimes\mathbb{Z}_2)_R}{O(N-1)_L\otimes(\mathbb{Z}_2)_R\otimes(\mathbb{Z}_2)_D}\approx
    \frac{SO(N)}{SO(N-1)}\otimes\mathbb{Z}_2,
\end{equation}
for $N\geq2$, and $\mathbb{Z}_2\otimes\mathbb{Z}_2$ for $N=1$, that is equivalent to the case $w=0$. Among the discrete subgroups, the third one (relating to a change in the sign of one vector) is only spontaneously broken. The second subgroup breaking is compensated by rotations $O(N)_L $, that leads to the appearance of the unbroken diagonal group $(\mathbb{Z}_2)_D$.

If one considers a weakly fluctuating configuration in the form $\phi(x)=\phi_0+\alpha(x)+\beta(x)$, $\psi(x)=\psi_0+\alpha(x)-\beta(x)$, then one finds the following mass spectrum of excitations
\begin{equation}
\begin{array}{lr}
m_{\alpha_1}=8\kappa^2(u+v+w), & m_{\alpha_i}=0,\\
m_{\beta_1}=8\kappa^2(u-v-w), & m_{\beta_i}=-8\kappa^2w,
\end{array}
\label{fluct1}
\end{equation}
with $i=2,\ldots,N$. The $N-1$ massless modes are Goldstone modes corresponding to the breaking of the continuous $SO(N)/SO(N-1)$ symmetry.

The submanifold $v=u-w$ is special. It corresponds to the sinusoidal phase of the model (\ref{ONxOM}) considered in \cite{Garel76,Bak76,Brazovsky76,Dzyaloshinsky77,Barak82,Kawamura88}. Wherein, a length of the vectors $\phi$ and $\psi$ remains undefined, but the sum of their length square is determine from the minimum conditions. The mode $\beta_1$ becoming massless corresponds to the continuous symmetry associated with $SO(2)_{\beta_1}$ rotations of the 2-vector $(|\phi|,|\psi|)$. Also, when $v=u-w$ and $|\phi|=|\psi|$, the group $(\mathbb{Z}_2\otimes\mathbb{Z}_2\otimes\mathbb{Z}_2)_R$ enlarges to $O(2)_R$. But this enlargement does not affect on the ground state degeneracy.

\subsection{$w=0$}

In this case, the minimum conditions (\ref{Z2xON}) does not determine the relative orientation of the vectors $\phi$ and $\psi$. Therefore, the broken symmetry is
\begin{equation}
\frac{SO(N)}{SO(N-1)}\oplus\frac{SO(N)}{SO(N-1)},\quad N\geq2,
\label{multisym}
\end{equation}
and $\mathbb{Z}_2\oplus\mathbb{Z}_2\equiv\mathbb{Z}_2\otimes\mathbb{Z}_2$ for $N=1$. There is $2N-2$ massless modes $\alpha_i$ and $\beta_i$ (see (\ref{fluct1})) in this case. When $u=v$, the symmetry enlarges to $O(2N)$ group with $2N-1$ Goldstone modes.

The symmetry (\ref{multisym}) is broken only when a multicritical point is tetracritical $u>|v|$ \cite{Fisher73}. A bicritical point $u<v$ describes symmetry breaking in only one order parameter, and another parameter remains zero ($u<-v$ is out of the stability region).

\subsection{$w>0$}

A planar (canted) ordering appears in this case
\begin{equation}
    \phi_0^2=\psi_0^2=\frac{-r}{2(u+v)}\equiv \kappa^2,\quad \phi_0\perp\psi_0,
\end{equation}
with the stability region
\begin{equation}
    u>0, \quad w>0, \quad u+v>0, \quad u-v>0.
\end{equation}
Using the symmetry of the GLW functional, the ground state can take the form $\phi_0=(\kappa,0,\ldots,0)$ and $\psi_0=(0,\kappa,0,\ldots,0)$. The group, right acting to the order parameter $\Phi$, is compensated entirely by $O(N)_L$ rotations. Thus the spontaneously broken symmetry is
\begin{equation}
    \frac{O(N)_L\otimes(\mathbb{Z}_2\otimes\mathbb{Z}_2\otimes\mathbb{Z}_2)_R}{O(N-2)_L\otimes(\mathbb{Z}_2\otimes\mathbb{Z}_2\otimes\mathbb{Z}_2)_D}\approx
    \frac{SO(N)}{SO(N-2)},
\end{equation}
for $N\geq3$, and $SO(2)\otimes\mathbb{Z}_2$ for $N=2$.

It is useful to choose a weak fluctuating field configuration in the form $\phi(x)=\phi_0+\alpha(x)+\beta(x)$, $\psi(x)=\psi_0+\tilde\alpha(x)-\tilde\beta(x)$, where $\tilde\alpha=(\alpha_2,\alpha_1,\alpha_3,\ldots,\alpha_N)$ and $\tilde\beta=(\beta_2,\beta_1,\beta_3,\ldots,\beta_N)$. Now, mass spectrum of excitations is
\begin{equation}
\begin{array}{lcr}
m_{\alpha_1}=8\kappa^2(u+v), & m_{\alpha_2}=8\kappa^2w, & m_{\alpha_i}=0,\\
m_{\beta_1}=8\kappa^2(u-v), & m_{\beta_2}=0, & m_{\beta_i}=0,
\label{fluct2}
\end{array}
\end{equation}
where $i\geq3$. Thus, we have $2N-3$ Goldstone modes. At $v=u-w$, the group $(\mathbb{Z}_2\otimes\mathbb{Z}_2\otimes\mathbb{Z}_2)_R$ enlarges to $O(2)_R$ again, but the spontaneously broken symmetry remains the same.

The submanifold $v=u$ is special, when a length of the vectors $\phi$ and $\psi$ is undefined as in the case $w<0$. The symmetry $SO(2)_{\beta_1}$ becomes broken spontaneously but not evidently. The additional Goldstone mode $\beta_1$ in (\ref{fluct2}) corresponds to the sliding degree of freedom of the spin-density wave. Similarly to the case $w<0$, this special submanifold demarcates the region of the ground state stability with the region where the minimum corresponds to only one non-zero order parameter.

\section{RG analysis}
\label{sect4}

In the one-loop approximation, beta-functions of the coupling constants are following
\begin{eqnarray}
\beta_u = -\epsilon u + \frac12\left(u^2(N+8)+v^2N+2vw+w^2\right),\nonumber\\
\beta_v = -\epsilon v + \frac12\left(uv(2N+4)+4v^2+2uw+w^2\right),\\
\beta_w = -\epsilon w + \frac12\left(w^2(N+2)+4uw+8vw\right).\nonumber
\end{eqnarray}
This system of equations predicts existence of eight fixed points (FP). Six of them turn to be well-known in the context of the models (\ref{On+Om}) and (\ref{ONxOM}). Each of them belongs to at least one of three submanifolds $w=0$ (the model (\ref{On+Om})), $v=u-w$ (the model (\ref{ONxOM})), and $v=u$. The later flat may describe some physically interesting model, but its interpretation is unknown for the author.
\begin{itemize}
\item Gaussian FP.
\item Heisenberg FP. This point falls on the line $u=v$, $w=0$, and corresponds to the $O(2N)$-model. Together with the GFP, it belongs to the all of three submanifolds.
\item Decoupled FP. It falls on the line $v=w=0$ and describes two decoupled $O(N)$-models. This multicritical point is always tetracritical.
\item Biconical FP. It is non-trivial point on the submanifold $w=0$. Depending on $N$, it can be tetracritical as well as bicritical. It describes two interacting $O(N)$-models. The submanifold $w=0$ is stable.
\item Chiral and antichiral FPs. These points belong to the submanifold $v=u-w$. They appear on the RG-diagram when $N$ is sufficient small in the sinusoidal phase. In this case they are marked as $S_{\pm}$. With $N$ increasing, they coincide at some $N_{c1}$ and become complex. With a further increase of $N$, these points appear again at some $N_{c2}$ and are marked as $C_{\pm}$. The chiral point $C_+$ describes a phase transition in the $O(N)\otimes O(2)$-model. Note that the submanifold $v=u-w$ is not fixed for the RG-equations. Nevertheless, this pair of the FPs belongs just to this submanifold for all values of $N$.
\item New FP $P_{1,2}$. These points belong to the stable submanifold $u=v$.
\end{itemize}

\begin{figure}[t]
\center
\includegraphics[scale=0.5]{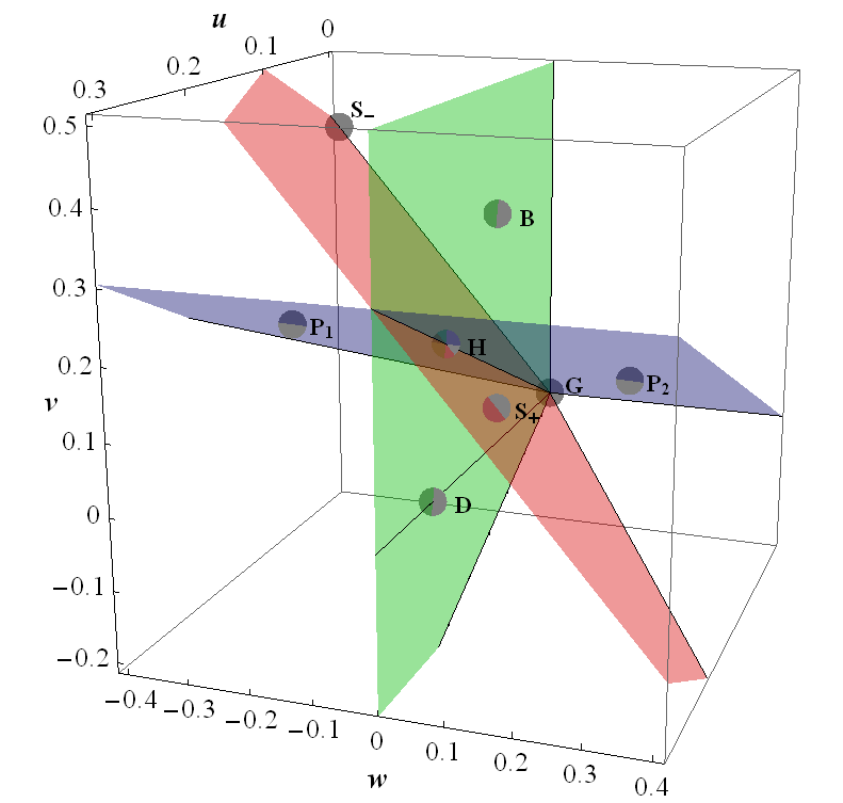}
\caption{\label{fig3}Qualitative position of the fixed points at $N=1$.}
\end{figure}%
Certainly, a position and stability of the FPs strongly depend on $N$. Qualitative diagram showing the position of the FPs in the physically interesting case $N=1$ is shown in fig. \ref{fig3}. Below, we consider evolution of RG-diagram with increasing of $N$. Course, the exact critical values of $N$, when a qualitative picture changes, require knowledge of higher orders in the $\epsilon$-expansion and resummation of the series. Fortunately, such a information obtained using different approaches is known for the models (\ref{On+Om}) and (\ref{ONxOM}). In addition, properties of the novel points $P_{1,2}$ as a function of $N$ is closely related to the properties of already studied points.

We find four critical values of $N$ associated with a coincidence of two or more FPs. All of these critical values of $N$ appear in the models (\ref{On+Om}) and/or (\ref{ONxOM}).

\begin{enumerate}
\item $N<N_H$.

One observes the stable fixed point is Heisenberg FP (fig. \ref{fig3}). This point is attractive in all of three models (\ref{ONxOM}), (\ref{On+Om}) and (\ref{Z2xON}) (fig. \ref{fig4}). There are two FPs $S_-$ and $P_1$ in the interesting region $w<0$, but both are saddle points. In the one-loop approximation $N_H=2$, but higher orders predict the value $N_H\simeq1.45$ \cite{Sokolov95,Pelissetto01,Calabrese04,Sokolov94,Pelissetto01-2,Pelissetto01-3,Calabrese02,Calabrese03}. At $N=N_H$, four points (namely $H$, $B$, $S_+$ and $P_1$) coincide. Since $N=1<N_H$, the Heisenberg fixed point with $O(2)$ symmetry corresponds to a second-order phase transition in the $N=1$ models (\ref{On+Om}) and (\ref{Z2xON}).
\begin{figure}[t]
\includegraphics[scale=0.42]{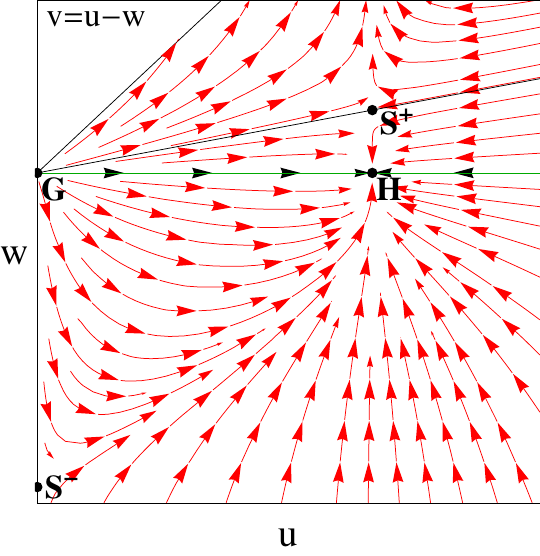}
\includegraphics[scale=0.42]{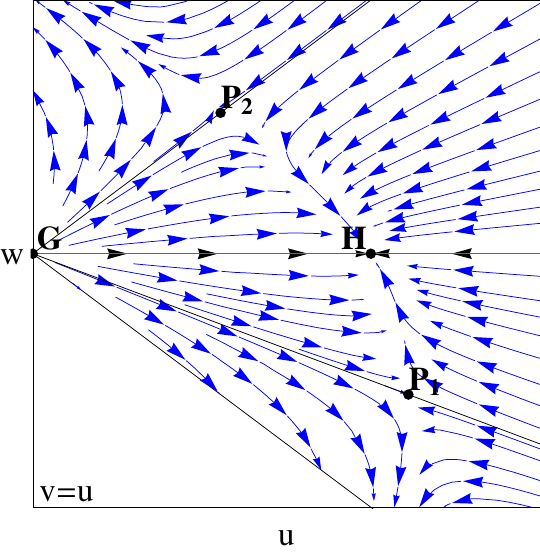}
\caption{\label{fig4}RG-flow on the submanifolds $v=u-w$ and $v=u$ at $N=1$.}
\end{figure}%

\item $N_H<N<N_{c1}$.

There are no attractive FPs in this case. Above the value $N_H$, the points $S_+$ and $P_1$ change the sign of their $w$-coordinate. So now, the points $S_+$ and $S_-$ fall in the region $w<0$. The point $B$ becomes tetracritical and stable in the model (\ref{On+Om}). When $N$ reaches to the value $N_{c1}$, the points $S_+$ and $S_-$ coincide and become complex. In the one-loop approximation, $N_{c1}\simeq2.20$, but $N_{c1}\simeq1.97$ in higher orders in $\epsilon$ \cite{Sokolov95,Pelissetto01,Calabrese04}.

\item $N_{c1}<N_H<N_D$.

As long as two points are complex-valued, just six fixed points are presented in the RG-diagram, but an attractive FP absents again, as well as FPs absent in the interesting region $w<0$. At $N=N_D$, two coincidence events occur, the point $P_1$ coincides with $P_2$, and the point $D$ with $B$. $N_D=4$ in the one-loop approximation, and $N_D\simeq2$ in higher orders.

\item $N_D<N<N_{c2}$.

\begin{figure}[t]
\center
\includegraphics[scale=0.5]{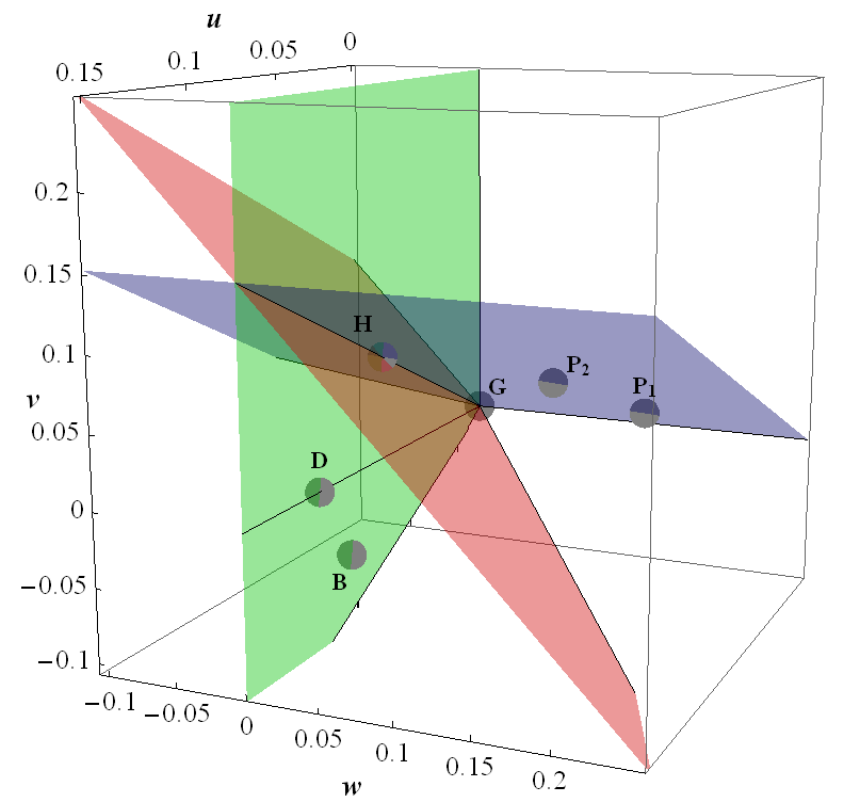}
\caption{\label{fig5}Qualitative position of the fixed points at $N=2,\,3$.}
\end{figure}%
Still, one observes six FPs (fig. \ref{fig5}), without attractive one and FPs in the region $w<0$. This case describes a situation in the physically interesting cases $N=2,\,3$.

\item $N_{c2}<N$.

At $N=N_{c2}$, the points $C_+$ and $C_-$ appear in the region $w>0$ of the RG-diagram. The first of them is stable. It describes a phase transition in the $O(N)\otimes O(2)$ model. The value $N_{c2}\simeq6$ \cite{Delamotte04,Sokolov95,Pelissetto01,Calabrese04,Sokolov94,Pelissetto01-2,Pelissetto01-3,Calabrese02,Calabrese03} (the one-loop result is $N_{c2}\simeq21.8$).
\end{enumerate}

Summarizing, we note that a stable (attractive) fixed point is present in the RG-diagram at $N<N_H<2$ and $N>N_{c2}$ but located in the region $w\geq 0$. Thus, a phase transition from the $\mathbb{Z}_2\otimes O(N)$ universality class is of first order for $N\geq2$.

\section{Monte Carlo results}
\label{sect5}

\begin{figure}[t]
\center
\includegraphics[scale=0.45]{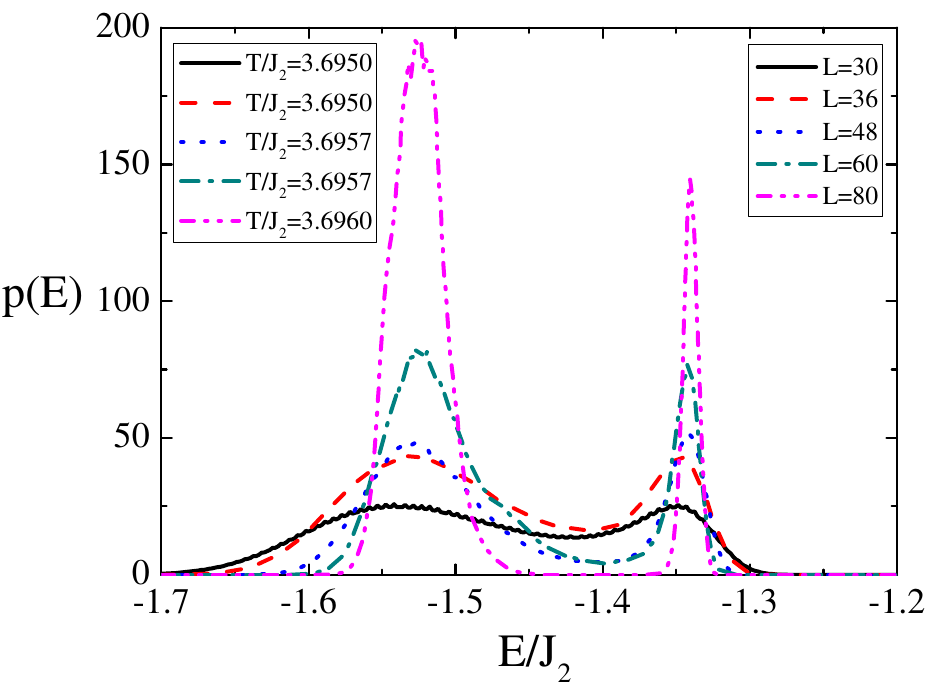}
\caption{\label{fig6}Energy distribution near the transition temperature for the $N=1$ stacked-J$_1$-J$_2$ model with $J_2/J_1=2/3$.}
\end{figure}%
Let's return to discuss the lattice models (\ref{model1}) and (\ref{model2}). To determine the order of a transition, we perform Monte Carlo simulation based on the Wang-Landau flat-histogram algorithm \cite{Wang01}. As long as we expect that a phase transition is of weak first order, we should consider large lattices, where the internal heat of a first-order transition becomes explicit. And the Wang-Landau algorithm has already proven itself to be sufficiently effective for such tasks \cite{Diep08,Diep08-2}.

The Wang-Landau algorithm \cite{Wang01} estimates accurately the density of states, $\rho(E)=e^{g(E)}$, which is defined as the number of spin configurations for any given $E$. The algorithm starts with a random lattice configuration, an empty array of the logarithmical energy density histogram $\rho(E)$, the empty visitation histogram $h(E)$, and some initial value (usually 1) of the weight constant $a$. Then, one chooses randomly a spin and its new orientation. A new spin configuration is accepted with the probability $e^{g(E_\mathrm{new})-g(E_\mathrm{old})}$, the element $h(E)$ of the visitation number histogram is increased by 1, and $g(E)$ is increased by $a$. This procedure is repeated until the visitation histogram is relatively flat, $|h(E)-\bar h|>0.8\bar h$ for each $E$. Then, the value of the weight constant $a$ is divided by $e$, the visitation number histogram is emptied, and the next step of the algorithm begins. We perform 30 such steps.

We consider $N=1,\,2$ and 3 for the both lattice models. For the model on a body-centered cubic lattice, we set $J_2=1$, $J_1=1.4$ and $1$. For the stacked-J$_1$-J$_2$ model on a simple cubic lattice, we set $J_2=1$, $J_1=1.5$ and $1$. The numerical estimation of the critical temperature is shown in table \ref{tab}.
{\begin{table*}[t]
\small
\center
\caption{Critical temperature $T_c/J_2$ in the model on a body-centered cubic lattice (BCC) and in the stacked-J$_1$-J$_2$ model (s-J$_1$-J$_2$).}
\begin{tabular}{c|c|c|c|c}
\hline
\hline
Model     &  $J_1/J_2$   &  $N=1$   &  $N=2$   &  $N=3$ \\
\hline
BCC       &   1.4        & 3.496(1) & 1.704(1) & 1.122(1)\\
\cline{2-5}
          &   1          & 4.094(1) & 2.004(1) & 1.317(1)\\
\hline
s-J$_1$-J$_2$ & 1.5      & 3.696(1) & 1.786(1) & 1.167(1)\\
\cline{2-5}
          &   1          & 4.173(1) & 2.034(1) & 1.331(1)\\
\hline
\hline
\end{tabular}
\label{tab}
\end{table*}%
}

\begin{figure}[t]
\center
\includegraphics[scale=0.45]{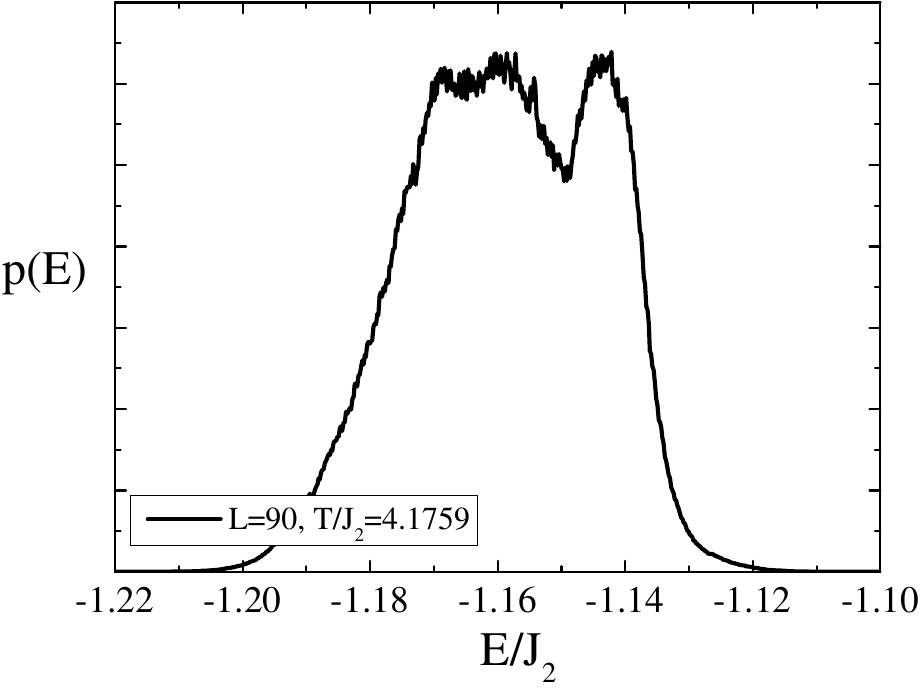}
\caption{\label{fig7}Energy distribution near the transition temperature for the $N=1$ stacked-J$_1$-J$_2$ model with $J_2/J_1=1$.}
\end{figure}

\begin{figure}[t]
\center
\includegraphics[scale=0.45]{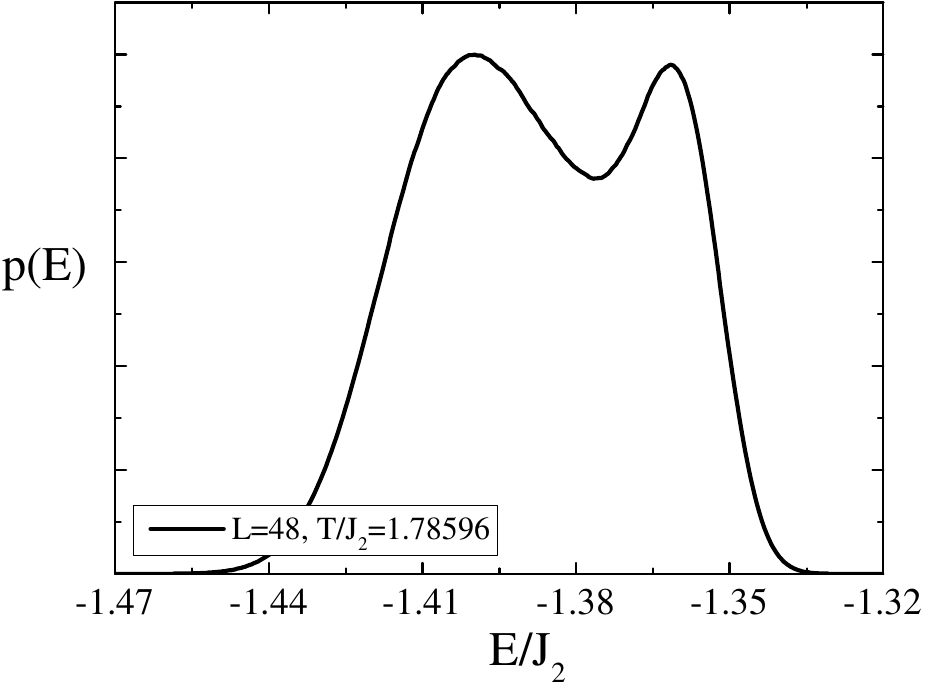}
\caption{\label{fig8}Energy distribution near the transition temperature for the $N=2$ stacked-J$_1$-J$_2$ model with $J_2/J_1=2/3$.}
\end{figure}

\begin{figure}[t]
\center
\includegraphics[scale=0.45]{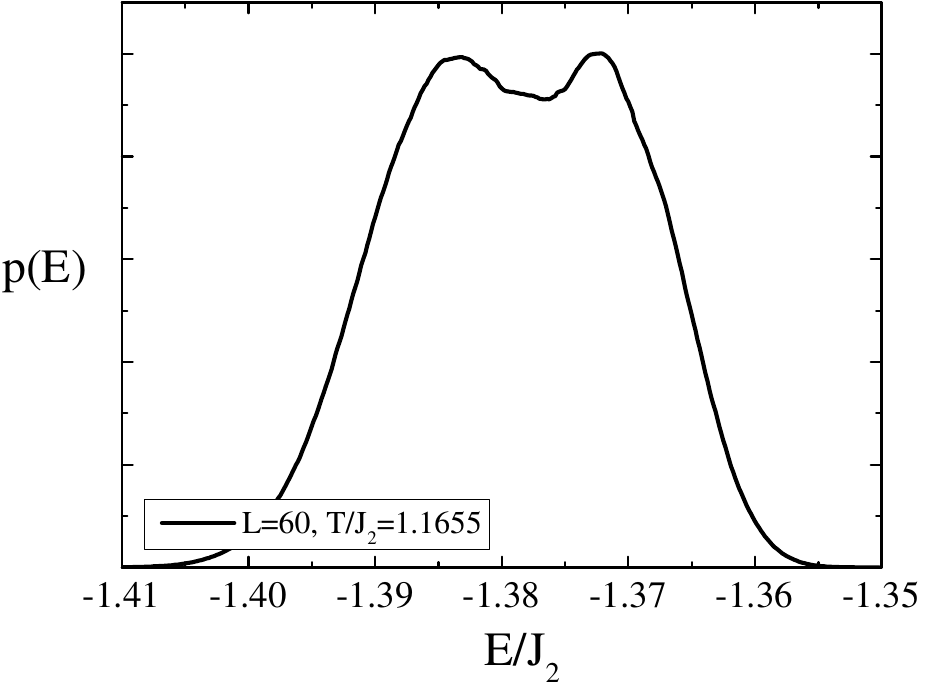}
\caption{\label{fig9}Energy distribution at the transition temperature for the $N=3$ stacked-J$_1$-J$_2$ model with $J_2/J_1=2/3$.}
\end{figure}

Note that the first order of a transition is weaker for the stacked-J$_1$-J$_2$ model, so one should take larger size of a lattice to determine the transition order. We show in figs. \ref{fig6}-\ref{fig9} the energy distribution only for the stacked-J$_1$-J$_2$ model.

Fig \ref{fig6} shows the evident first order transition for the case $N=1$ and $J_2/J_1=2/3$. But the internal heat of the transition does not demonstrate a dependance on the lattice size $L$. It excludes the pseudo-first order behavior observed in the J$_1$-J$_2$ model on a square lattice (in two dimensions) \cite{Jin12,Jin13,Kalz12}.

For the case $J_2/J_1=1$, the first order of the transition is less evident. In \cite{Murtazaev15,Murtazaev15-2}, it has been concluded that the transition is of second order. But the critical exponents differ from the exponents of the $O(2)$ model (e.g., $\nu\approx0.671$). We estimate the index as $\nu\approx0.54(1)$ in the both models at $J_2/J_1=1$ and $N=1$. The same estimation of $\nu\approx0.55$ has been found in \cite{Murtazaev15-2}. Such a value of the exponent $\nu$ is close to the mean-field value for the tricritical behavior, so one may assume that the transition corresponds to a tricritical point, and at $J_2/J_1>1$ a transition is of second order. However, we find the first order transition considering large lattices $L\leq90$ (see fig. \ref{fig7}).

For the cases $N=2,\,3$, the evidence of the first order of a transition becomes less (see figs. \ref{fig8}, \ref{fig9}). Thus, for the case $J_2/J_1=1$, one should consider the lattice size $L\geq120$. Nevertheless, the first order of a transition is observed in all considered cases.

\section{Conclusion}

We performed RG-analysis of the $\mathbb{Z}_2\otimes O(N)$ model describing in particular the critical behavior in the class of frustrated antiferromagnets with a collinear spin ordering and an additional twofold degeneracy of the ground state. In the case $N=1$ interesting also in the context of the Ashkin-Teller model, one expects that a phase transition with the $\mathbb{Z}_2\otimes \mathbb{Z}_2$ symmetry breaking is of second order from the universality class of the $O(2)$ model or of first order dependently on  initial values of the coupling constants. In addition, crossover tricritical exponents may be observed, associated with the fixed point $P_1$ belonging to a submanifold, which separates these two types of the critical behavior. In the stacked-J$_1$-J$_2$ model on a simple cubic lattice and J$_1$-J$_2$ model on a body-centered cubic lattice, the situation with a second-order phase transition is not realized.

For $N\geq2$, a first order transition is predicted for the $\mathbb{Z}_2\otimes O(N)$ universality class. At $N=2$, this class is equivalent to the symmetry class of the $O(N)\otimes O(2)$ model (\ref{ONxOM}) corresponding to magnets with a planar spin ordering, where $\mathbb{Z}_2\otimes SO(2)$ symmetry is broken. In this class, a transition must be of a first order \cite{Sokolov95,Pelissetto01,Calabrese04,Zumbach93,Delamotte00}. At the same time, one discusses a possibility that this transition is of weak first order or almost second order. This is intended to explain the pseudo-scaling and pseudo-universality observed for this symmetry class (see \cite{Delamotte04} for a review). In terms of the renormalization group, an imitation of a second order transition is possible, if the RG diagram contains a sufficient small region attractive for RG trajectories starting from a quite wide range of initial parameters, and where the RG-flow is rather slow. The existence of such a region in the $O(2)\otimes O(2)$ model has been studied in works \cite{Zumbach93,Delamotte00}. Of cause, this region has $w>0$. An almost second order transition is possible in the $\mathbb{Z}_2\otimes O(N)$ model. But here, a region of slow RG-flow must be in the region $w<0$, but in this case, a non-trivial local minimum of the RG-flow is not found.

A region with a slow RG-flow may exist if coordinates of some fixed point are complex-valued but with small imaginary part. Such a situation is observed in the $O(N)\otimes O(2)$ model (\ref{ONxOM}) with $N=3$ \cite{Zumbach93,Delamotte00}. The points $C_\pm$ have the real part of their coordinates close to the submanifold $v=u-w$ and $w>0$. But such points absent in the $\mathbb{Z}_2\otimes O(N)$ model (\ref{Z2xON}) for $N\geq2$. So in the region $w<0$, a slow RG-flow region is possible only if $|w|$ is sufficient small.

We guess that in the $\mathbb{Z}_2\otimes O(N)$ model the pseudo-scaling behavior at the almost second phase transition has another origin. Lets consider the large $N$ limit. In the ordered phase, $\psi=\sigma\phi$, where $\sigma=\pm1$. To make the large multiplier $N$ explicit, one needs to make replacements $u\to\frac{u}{N}$, $v\to\frac{v}{N}$ and $w\to\frac{w}{N}$. But then $|\phi|^2=N\kappa^2$ (see (\ref{absphi})), and the GLW functional (\ref{Z2xON}) is
\begin{eqnarray}
    F=&&\int d^3x\,N\left((\partial_\mu\sigma)^2+(r+2v+2w)\sigma^2+u\sigma^4
    \right.\nonumber\\
    &&\left.+\mathcal{O}\left(\frac1N\right)\right).
\end{eqnarray}
Thus, we expect that a weak first-order transition in the $\mathbb{Z}_2\otimes O(N)$ model and corresponding lattice models has the pseudo-scaling behavior from the universality class of the three-dimensional Ising model ($\nu\approx0.63$) for large values of $N$ as well as large values of $J_2/J_1$.
\medskip

This work is supported by the RFBR grant No 14-02-31448 and No 16-32-60143.

\end{document}